\documentclass[USenglish,oneside,twocolumn]{article}

\usepackage[utf8]{inputenc}

\usepackage[big]{dgruyter_NEW}
\usepackage{subfiles}

\usepackage{natbib}
\setcitestyle{numbers,square,comma}

\usepackage{eurosym}
\usepackage{tablefootnote}
\usepackage{listings}
\usepackage[group-separator={,}]{siunitx}

\begin{document}
  \author*[1]{Matthew Smith}
  \author[2]{Daniel Moser}
  \author[3]{Martin Strohmeier}
  \author[4]{Vincent Lenders}
  \author[5]{Ivan Martinovic}
  \affil[1]{Department of Computer Science, University of Oxford, United Kingdom, E-mail: matthew.smith@cs.ox.ac.uk}
  \affil[2]{Department of Computer Science, Z\"{u}rich, Switzerland, E-mail: first.last@inf.ethz.ch}

  \affil[3]{Department of Computer Science, University of Oxford, United Kingdom, E-mail: martin.strohmeier@cs.ox.ac.uk}
  \affil[4]{armasuisse, Switzerland, E-mail: first.last@armasuisse.ch}
  \affil[5]{Department of Computer Science, University of Oxford, United Kingdom, E-mail: ivan.martinovic@cs.ox.ac.uk}
 
    \title{\huge Analyzing Privacy Breaches in the Aircraft Communications Addressing and Reporting System (ACARS)}
  \runningtitle{Analyzing Privacy Breaches in the Aircraft Communications Addressing and Reporting System (ACARS)}
\begin{abstract}
{The manner in which Aircraft Communications, Addressing and Reporting System (ACARS) is being used has significantly changed over time. Whilst originally used by commercial airliners to track their flights and provide automated timekeeping on crew, today it serves as a multi-purpose air-ground data link for many aviation stakeholders including private jet owners, state actors and military. Since ACARS messages are still mostly sent in the clear over a wireless channel, any sensitive information sent with ACARS can potentially lead to a privacy breach for users. Naturally, different stakeholders consider different types of data sensitive. In this paper we propose a privacy framework matching aviation stakeholders to a range of sensitive information types and assess the impact for each. Based on more than one million ACARS messages, collected over several months, we then demonstrate that current ACARS usage systematically breaches privacy for all stakeholder groups. We further support our findings with a number of cases of significant privacy issues for each group and analyze the impact of such leaks. While it is well-known that ACARS messages are susceptible to eavesdropping attacks, this work is the first to quantify the extent and impact of privacy leakage in the real world for the relevant aviation stakeholders.}
\end{abstract}

\startpage{1}

\maketitle
\section{Introduction}
\label{sec-intro}

Recently, potential privacy issues of aviation data became apparent through a series of high-profile cases. Both criminals and journalists, amongst others, exploited this type of data. Of the former, financial traders used aircraft movements to track potential deals to later use for insider trading~\cite{Daniels2012}, in others flights relating to alleged gold smuggling were hidden~\cite{Grabell2011}. An example of the latter comes in the movements at Geneva airport identifying regular flights relating to Teodoro Obiang Nguema Mbasogo, the incumbent President of Equitorial Guinea, wanted in Europe on charges of embezzlement~\cite{Toor2016, Pilet2016a}. This led to the creation of a Twitter account, GVA Dictator Alert, reporting on aircraft belonging to alleged dictators arriving at Geneva airport~\cite{Pilet2016a}. Privacy in aviation is rapidly becoming hard to achieve and is being further hindered by outdated, insecure systems.

As with many areas of transport, aviation strives for becoming more `connected', utilizing modern data link technologies to improve efficiency. However, data link usage is not new to aviation; since the 1970's, links of some form have been used to transfer data to aircraft. A prime example is the Aircraft Communications, Addressing and Reporting System (ACARS), which has evolved from being a single purpose tool primarily for commercial aviation to being widely deployed on many types of aircraft.

ACARS provides data link communications between aircraft and entities on the ground and is used for many different purposes, from air traffic control (ATC) to management of aircraft fleet. Using a basic ASCII character set, a range of messaging formats and protocols are defined to provide services. However, the system was not designed with security in mind, with solutions being offered much later as add-ons. This means that the vast majority of ACARS traffic is in the clear. With a variety of stakeholders each with differing requirements, it has gradually begun to be used for purposes it is not equipped to handle. 

One way this manifests itself is in leaking a significant amount of private data. Some messages overtly leak sensitive information---particularly in the case of commercial aircraft, efforts to transfer data to make the passenger experience better often breach privacy. We observed instances of medical information and full credit card details being transferred via ACARS. On top of this, some messages which initially appear innocuous can also leak significant amounts of information. For stakeholders who wish to, e.g., hide flight information, ACARS messages are regularly used to report position or their destinations and route thus undermining this effort. 

Indeed, concerns about privacy in ACARS have been highlighted by a number of unofficial sources. As far back as 1998, discussion on the clear-text nature of avionic data link highlighted the leakage of medical and passenger information~\cite{Wolper1998}. More recently, an ex-pilot discussed his view of ACARS usage from the cockpit, acknowledging the eavesdropping threat and providing anecdotes of messages being circulated widely amongst ground networks despite intended for a narrow group of people~\cite{Pascoe2015}.


\vspace{-0.2cm}
\subsection*{Contributions}
Due to the fact that ACARS is used in many ways by different users, systematically analyzing such data would require access to a great deal of insider information. However, simply providing anecdotal evidence is not sufficient to illustrate the severity of this problem. As such, our contributions in this paper are as follows:

\begin{itemize}
\item We provide a taxonomy of data link users and discuss the privacy expectations of each group.
\item In order to demonstrate the privacy issues to each group, we provide evidence from our data collection.
\item We demonstrate why the information leaked for each group is a privacy breach.
\item We discuss why such privacy breaches occur, despite ACARS being known to be an unsecured data link.
\end{itemize}

The remainder of this work is structured as follows: Section~\ref{sec-threat} introduces the threat model that privacy-conscious actors in aviation face. Section~\ref{sec-background} describes the necessary background on aviation and its main data link ACARS whilst Section~\ref{sec-provisions} analyzes the existing privacy provisions available to aviation users. Section~\ref{sec-framework} presents our novel framework for aviation privacy comprising concerned stakeholders and privacy requirements. Section~\ref{sec-collect} discusses our data sources, Section~\ref{sec-expectations} identifies stakeholder privacy requirements whilst Section~\ref{sec-breach} evaluates how privacy requirements are being breached on the ACARS data link. This is followed by an analysis of the implications in Section~\ref{sec-discussion}. Finally, Section~\ref{sec-related} discusses the related work before Section~\ref{sec-conclude} concludes.

\vspace{-0.1cm}
\section{Threat Model}
\label{sec-threat}

To frame our discussion of the privacy landscape in aviation, we first propose the typical threat to users of avionic data link systems. We consider our attacker to be passive with respect to the medium, only ever receiving messages. 

Due to the general lack of security in ACARS the barrier-to-entry for an attacker is low. We presume an attacker to be moderately resourced, having access to standard desktop computers, commodity software-defined radios (SDRs) and antenna. We further presume the attacker has a moderate level of technical capability, i.e. can set up and use the equipment, with the ability to produce tools to operate the SDR. Given the range of uses for avionic data link, different attackers are thus likely to have vary in intention. Primarily, we model an attacker collecting data for either criminal gain, to achieve competitive advantage or for some kind of surveillance. 

Attackers seeking criminal gain might focus on financial or operational data allowing them to steal assets or blackmail victims. Those looking for competitive advantage might seek to track business aircraft in order to predict business actions, as alluded to in Section~\ref{sec-intro}. Threat actors aiming to use data link for surveillance might simply want to know if an aircraft is in the air, others might want to acquire more detailed information about its location and status. Due to the range of possible intentions of threat actors, we note that an attacker might deploy sensors in multiple locations or collect in a mobile fashion.


\vspace{-0.1cm}
\section{Background}
\label{sec-background}

In this section we introduce the key concepts needed to contextualize the problem of sensitive data leakage from ACARS. We first explain the aviation scenario before looking at ACARS and how it is used in detail. 
\vspace{-0.2cm}
\subsection{Aviation}

Airspaces are complicated, safety-critical environments which rely on quick, accurate communication between all parties involved. Each country or region has Air Navigation Service Providers (ANSPs) which administer ATC for a given region. It is their responsibility to ensure that the aircraft in the airspace have sufficient separation, to allocate take-off and landing slots, and to handle any emergency situations which may arise. In order to do this, voice and data communications are used extensively both on the ground and in the air. Furthermore, aircraft are often part of a fleet of numerous airframes. Monitoring these aircraft for information such as location, estimated arrival times and maintenance data allows for faster turnarounds and more efficient operation. 

A number of systems are used to manage the civil airspace with future developments focussing on data link rather than voice communications. This involves introducing systems such as Automatic Dependent Surveillance-Broadcast (ADS-B) for tracking aircraft~\cite{FederalAviationAdministration2016}, making greater use of systems such as Mode S to provide situational awareness (see ~\cite{Schafer2016}) as well as utilizing existing systems such as ACARS until new ones are deployed.

Broadly, aviation can be split into flights falling into---and outside of---the category of \textit{general aviation}. This category is defined as all scheduled air services or non-scheduled air transport operations for renumeration or hire~\cite{InternationalCivilAviationOrganizationICAO2009} though this divide is  somewhat simplistic and we expand upon it in Section~\ref{sec-framework}. However, the divide can still heavily influence the usage of avionic systems.

\vspace{-0.2cm}
\subsection{Aircraft Communications Addressing and Reporting System}
\begin{figure}
\centering
\includegraphics[scale=0.25]{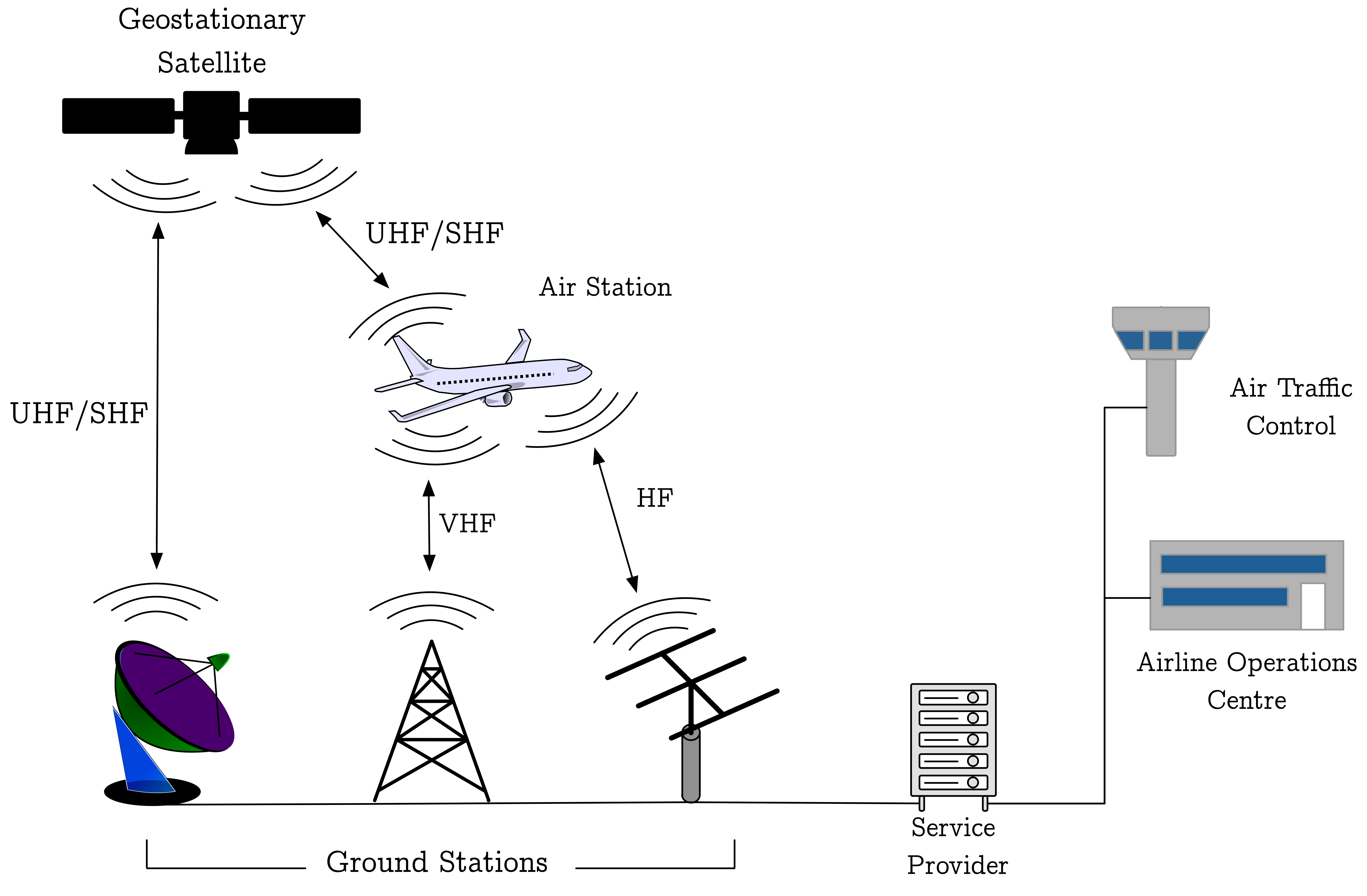}
\caption{Representation of the ACARS sub-systems}
\label{fig-acars-sys}
\end{figure}

Originating from the VHF network created in 1978~\cite{Oishi2015a}, ACARS provides an ASCII character-based data link between aircraft and ground stations; it is defined by ARINC 724B-6~\cite{AeronauticalRadioInc.ARINC2012}. Originally specified for Very High Frequency (VHF), High Frequency (HF) and Satellite Communications (SATCOM) links were added, allowing worldwide coverage. Typically, VHF is used over populated land and small bodies of water, SATCOM extends coverage to oceanic and rural land areas, with HF providing worldwide coverage. Furthermore, VHF is offered through older technology known as Plain Old ACARS (POA) and the newer VHF Data Link mode 2 (VDLm2), which has a higher data rate. A representation of the ACARS subsystems can be seen in Figure~\ref{fig-acars-sys}. An aircraft will select one of the three communication methods based on signal strength---typically, the priority from high-to-low is VHF, SATCOM and HF. 

\begin{figure}
\centering
\includegraphics[scale=0.4]{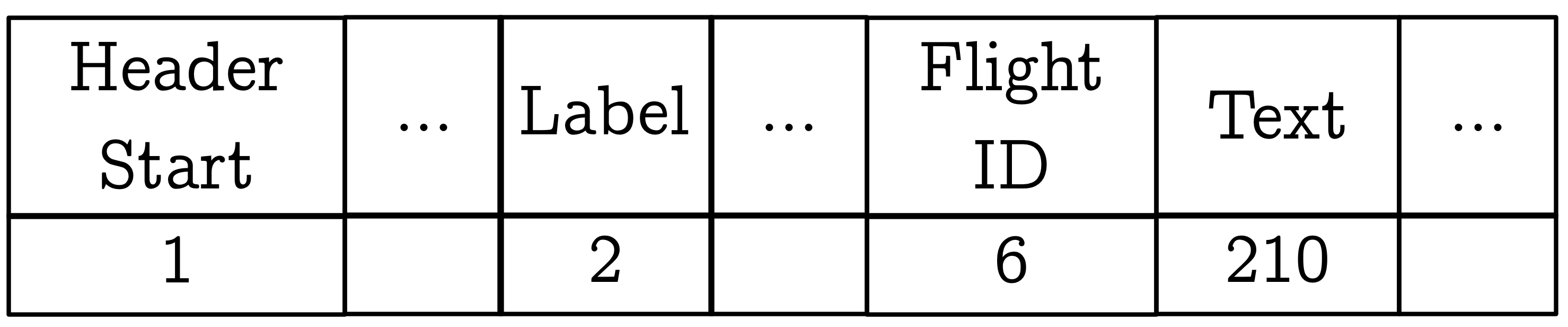}
\caption{Summarized ACARS message format, identifying the relevant fields for this paper}
\label{fig-mesg-format}
\end{figure}

ACARS messages are mainly composed of a text field---a summarized version of the message format can be seen in Figure~\ref{fig-mesg-format}. Routing is achieved via the flight ID and label, being used by a Communications Management Unit (CMU) to send the message to the correct on-board system. Message content is structured based on the ARINC 620-8 standard~\cite{AeronauticalRadioInc.ARINC2014}. These messages cover a wide range of purposes including weather report requests, aircrew free-text messages or time requests. Indeed, messages of different purposes will be originated by different parties on board or on the ground.

\vspace{-0.2cm}
\subsection{Usage of ACARS}

Many aircraft systems can originate ACARS messages, creating a range of types of information sent over the link. Indeed, many of the services it provides now were not part of the original intention of the system---ACARS was initially designed to provide a logging system to monitor hours worked by crew~\cite{Oishi2015a}. It has now evolved significantly, providing services broadly divided into air traffic control and airline administrative/operational control.

ATC messages over ACARS provide safety-critical services such as negotiating clearances and communicating weather and aerodrome information. These services are particularly useful in congested airspace, where voice VHF frequencies are often under heavy load. In the case of negotiating a clearance to fly a particular route, an aircraft will use a predefined ACARS message exchange protocol in order make a request, whereas for weather and aerodrome information, a ground station will reply with data based on a request from an aircraft. In many ATC messages, an aircraft will reveal where it is flying to and sometimes where it currently is.

Airline administrative and operational control is not confined to airlines but to fleet managing in general. Examples include transferring flight plans, maintenance information, loading sheets or free-text crew communications with ground staff. Similar to ATC messages, flight plans can contain information about what the aircraft will do next, whereas many of the other of this category of message are relating to those onboard. Maintenance information messages are used to ensure that aircraft issues can be fixed upon landing whilst loading sheets are used to transfer what quantity of passengers and cargo are on board so pilots can configure the aircraft's systems. Most loosely defined are free-text messages which has been observed being used for anything from requesting aid for a passenger on landing to asking for sports scores or news updates. Generally, these messages are not immediately safety critical but instead can contain sensitive information with respect to the user of the aircraft. 

\vspace{-0.2cm}
\subsection{Securing ACARS}
\label{sec-securing}

Although ACARS has no security by default, some security solutions of varying complexity and effectiveness have been developed as optional `add-ons'. The most comprehensive systems are based on the ARINC 823P1 standard ACARS Message Security (AMS)~\cite{AeronauticalRadioInc.ARINC2007} and implementations based on this standard such as Secure ACARS~\cite{Roy2004}. These provide message confidentiality and authentication. Cryptography used in Secure ACARS matches the US National Security Agency's Commercial National Security Algorithm (CSNA) Suite though with older Suite B parameters~\cite{NationalSecurityAgency2015}, thus is of industry standard. However, little public work exists on the implementations as they are proprietary. Whilst no official usage figures have been published, we have not observed consistent usage of AMS on either SATCOM or VHF channels. 

Other proprietary solutions developed by individual airlines or manufacturers exist, some of which are highly insecure (such as monoalphabetic substitution ciphers~\cite{Smith2017}) and unsuitable in providing confidentiality or authentication. This is used somewhat widely across a range of stakeholder groups.


\vspace{-0.1cm}
\section{Protecting Privacy}
\label{sec-provisions}

A number of methods exist to protect privacy for data used in aviation, though not all are aviation specific. Furthermore, the ability to use each depends on the type of stakeholder. This section explains these provisions, as well as the realities and drawbacks of each.
\vspace{-0.2cm}
\subsection{Data Protection Legislation}
\label{sec-dataprot}
In many parts of the world, data protection legislation is a key measure in enabling citizens to protect their privacy. Whilst data protection provisions cannot guarantee privacy from the outset, they can at least provide a legal basis for defending it. This is particularly relevant to the aviation scenario; those on board the aircraft are unlikely to control how their data is treated, and the primary method of transferring data is by default not secured. 

As an example, the current European Union (EU) data protection regulation which requires member states to implement such legislation in their national laws was introduced in 1995~\cite{EuropeanParliament1995}. Section 8, Article 17 mandates the security of data  processing. Specifically, the data controller must ensure ``appropriate technical and organizational'' measures are taken to protect against loss, accidental disclosure and modification. Instances where inadequate security in conjunction with ACARS causes sensitive data to be transmitted in cleartext is an obvious breach of this regulation.
Unfortunately, this is a post-hoc solution which protects those affected by threat of legal action. As such this is powerful but slow-moving; the level and speed to which it is enforced may vary within the EU. This is set to become even more powerful in the coming years as new legislation affect all those processing data in the member states, as well as increasing fines to the greater of 4\% of company turnover or \euro 20 million~\cite{EuropeanParliament2016}.

\vspace{-0.2cm}
\subsection{Blocking of Public Aircraft Displays}
\label{sec-asdiblock}
One way non-commercial aircraft can restrict shared information is through government-level blocking programs. The most complete and public example is a block on the Aircraft Situation Display to Industry (ASDI) data feed, managed by the American Federal Aviation Administration (FAA)~\cite{FederalAviationAdministration2013}. The FAA allows organizations to subscribe to the ASDI feed, though obtaining access to this feed is not easily feasible.

Aircraft owners can apply to the FAA to have their information removed from the ASDI feed.
The most stringent block is at the FAA level, where flight information relating to a specific aircraft is removed before it is disseminated as part of the ASDI feed. Data can also be blocked at the subscriber level, who are then contractually obliged to not make information on blocked aircraft public. However, some information, usually existence, is available via flight trackers for an subscriber level block.

Until June 2011, obtaining such a block required an aircraft owner to demonstrate a `valid security concern', which amounts to a threat against a person, organization or the aircraft traveling to a location known to have a terrorist threat~\cite{FederalAviationAdministration2011}.  From this we can determine that those who used an ASDI block at this time had a definite, FAA approved cause for privacy. However the FAA relaxed this in December 2011~\cite{FederalAviationAdministration2011a}, now allowing any non-commercial aircraft to request a block~\cite{FederalAviationAdministration2013}. Even though a security concern is no longer required, requesting a block on an aircraft is still a definite attempt to achieve privacy. As such, we consider anything which undermines such a block to be a privacy breach.

\begin{table}[t!]
\centering
\small
\caption{Business aircraft using ACARS registered in off-shore financial centers (countries with more than 5 occurrences listed). }
\begin{tabular}{p{2.5cm}|p{4cm}}
OFC & Number of registered aircraft\\   \hline
Isle of Man        &118  \\
Malta        		&61   \\ 
Bermuda               &56   \\ 
Cayman Islands        &  49  \\ 
Aruba        &    22\\ 
UAE        &    17\\ 
Hong Kong        &    6\\ 

\end{tabular}
\label{tab-ofcs}
\end{table}

\vspace{-0.2cm}
\subsection{Obscuring of Public Registers}
\label{sec-obscuring}
Similarly to blocking of flight data, some stakeholders use third-party entities to register their aircraft to conceal the real owner from public records. Popular methods include the use of shell companies (often off-shore), special aircraft registration services, wealth management companies and trusts. We found over 12,000 aircraft registered by trusts and specialised services in the FAA records, accounting for over 3.75\% of all aircraft registrations in the US. Another 53,000 or 16.5\% are registered to Limited Liability Companies (LLC), many of which follow the naming scheme NXXXXX LLC, where the so-called N number is the registration of the aircraft. While careful mining of public records may unearth the real operators of these aircraft, the popularity of this type of public record obfuscation shows the importance of privacy provisions. 

Furthermore, 353 of the business aircraft in our collected ACARS data were registered to shell companies in recognized off-shore financial centers, with the most popular being the Isle of Man, Malta, Bermuda and the Cayman Islands (see Table \ref{tab-ofcs}).\footnote{We apply the definition of `off-shore' used by Rose and Spiegel~\cite{rose2007offshore}.} This means that more than 20\% of the business jets seen in our collection area in Central Europe are registered in such off-shore centers.

\vspace{-0.2cm}
\subsection{Use of ACARS Security Mechanisms}
\label{sec-security-mechanisms}

As described in Section~\ref{sec-securing}, the standardized approach to security in ACARS is AMS. This is the most direct way of protecting many aspects of privacy when using ACARS as it leaves very little of the message content in the clear. It should be noted that whilst other security solutions do exist, such as the substitution cipher described in~\cite{Smith2017}, they are of extremely variable quality and thus cannot be considered to protect privacy.

\vspace{-0.2cm}
\subsection{Radio Silence}
\label{sec-silence}

Arguably the most drastic option, radio silence involves an aircraft opting to not use an avionic technology such as ACARS. In some instances, this is the ability to turn the system on and off at will; e.g., military aircraft are permitted to turn ADS-B off for operational purposes~\cite{InternationalCivilAviationOrganization2013b}. While this protects privacy by greatly reducing information leakage, it severely degrades service. For ACARS this degradation is mostly operational and will cause inconvenience to ATC, aircraft operators and crew.

\vspace{-0.1cm}
\section{Aviation Privacy Framework}
\label{sec-framework}
In order to assess where privacy breaches occur, we propose a framework for privacy for the aviation scenario. We first introduce our taxonomy of four fundamental privacy concepts that are important in aviation. Secondly, we describe the relevant stakeholder categories, which are responsible for most of the data link usage in aviation. 
\vspace{-0.2cm}
\subsection{Privacy Concepts}
\label{sec-concepts}

To understand potential privacy breaches, we define four categories of aviation privacy: existence, intention, status, and passenger/cargo information. A given message may fit into multiple categories. Indeed, not all aircraft will transmit in each category and not all stakeholders will consider each to be sensitive.

\vspace{-0.5cm}
\subsubsection{Existence}
\vspace{-0.1cm}
The most basic concept of aviation privacy is existence, i.e., information relating to a particular aircraft's temporal and spacial whereabouts. Consequently, a breach of this concept requires the combination of both, knowledge about the operator or user of a given aircraft and acquisition of data on the aircraft's transmissions.

Preserving the privacy of existence is difficult as aircraft are required to be part of mostly public records and are easily spotted with the naked eye at airports and in the air. However, their use of a multitude of public communication channels during flight makes the breach of this concept automatically exploitable and most importantly scalable.

\vspace{-0.5cm}
\subsubsection{Intention}
\vspace{-0.1cm}
Determining an aircraft's origin and destination is less trivial than visually spotting it at airports. For commercial aviation, details of their routes are required to be public (see~\cite{FederalAviationAdministration2011}) but for all others, it is understandable why they might wish to hide their flight route. Receiving very few messages can fully reveal an aircraft's intentions. As an aircraft, e.g., negotiates ATC clearance it typically includes destination and route revealing not only where an aircraft is travelling, but also which path it takes. 

Similarly, communication revealing only the destination airport breaches intention privacy. Even when an aircraft operator avoids requesting airport clearance via ACARS, other seemingly innocent messages leak intention. An example are weather reports needed to safely navigate the airspace: an aircraft requesting weather information for an aeronautical waypoint or airport will be highly likely to visit the location it enquired about. 
\vspace{-0.5cm}
\subsubsection{Status}
\vspace{-0.1cm}
Since ACARS is used to report to operations and administrations teams, it can reveal a wide range of additional data relating to on board happenings beyond other surveillance mechanisms, such as the maintenance state of the aircraft and what it is doing right now. 

Status information can be revealed through systematic means, such as situational reporting. This includes position, altitude and speed, thus resulting in varying sensitivity for different stakeholder groups. Those which value intention and existence privacy are also likely to value protecting status. However, a significant portion of status messages are of a less predictable format. For example, if the cause of some fuel reports, maintenance requests and logbook entries caused an unexpected change in flight plan, this could contribute to a leak of sensitive status information.
\vspace{-0.5cm}
\subsubsection{Passenger/Cargo Information}
\vspace{-0.1cm}
For the concepts described above, most messages follow a regular structure and transmission pattern. In contrast, passenger and cargo information is found within free-text messages used by crew to connect with their airline operations centers. Communicating with the ground is operationally useful, e.g. when sending advance lists of passengers on a layover. However, when transferring such data in the clear, there are potential legal, ethical and financial implications. This information is inherently sensitive; at the most basic level, it ties passengers or aircraft contents to a specific departure and arrival airport. In more extreme cases, data such as medical condition or payment details could fall under this category. Indeed, passengers and cargo owners do not have agency over any data sent in this way. 

As most relevant content is transmitted in unstructured free-text messages, the concrete privacy issues under this concept vary significantly from airline to airline.

\vspace{-0.2cm}
\subsection{Aviation Stakeholders}
\label{sec-stakeholders}

The aviation scenario has a diverse set of stakeholders. In this section, we categorize these stakeholders along their respective objectives and needs, followed by an analysis of their privacy requirements. 

We build our categories based on the ICAO classification of civil aviation, which divides the group into commercial air transport and general aviation~\cite{InternationalCivilAviationOrganizationICAO2009}. Commercial air transport is defined as ``an aircraft operation involving the transport of passengers, cargo or mail for remuneration or hire'', with general aviation comprising all civil aviation outside of this definition. As such, state, business and non-operational military aviation fall under the general aviation category if the flights are not on hired aircraft and they are not on military missions. Table \ref{tab-actype} illustrates the breakdown of the different stakeholder types, on which we will elaborate in this section.

\vspace{-0.5cm}
\subsubsection{Business}
\vspace{-0.1cm}
Business stakeholders typically fly jets capable of 4--20 passengers. Gulfstream's G-range, Cessna's Citation jets and Bombardier's Learjet and Challenger aircraft are amongst the most popular choices. There are also business airliners based on commercial airframes produced by Boeing and Airbus, which in their VIP and corporate versions constitute the high-end of the market and are capable of carrying 100 passengers or more. According to ICAO definitions, business flights can either be commercial on-demand services (in the case of chartering an aircraft through a hire company) or, if the aircraft is owned by the operator and used without hire, can be counted under general aviation. They are used to transport business personnel to meetings, conferences or other gatherings and as such the place to which they travel can reveal a lot about the intentions of those on board. 

\vspace{-0.5cm}
\subsubsection{Commercial}
\vspace{-0.1cm}
A significant part of aviation is composed of commercial air transport, which we define as non-business commercial air transport. This can be a chartered or scheduled flight carrying passengers, cargo or mail. Usually these are large aircraft built by Airbus or Boeing, carrying hundreds of passengers or significant amounts of cargo. Passenger aircraft will often carry some cargo in the hold, cargo aircraft, on the other hand, do not tend to carry passengers. 

Commercial aircraft heavily use data link for ATC in congested airspace and fleet management, with the exact usage varying depending on the operator. It appears a significant portion of this communication is automated.

\vspace{-0.5cm}
\subsubsection{Military}
\vspace{-0.1cm}
We classify any aircraft operated by a national airforce as military aircraft. Flying many types of aircraft, military stakeholders operate in a very different way to commercial aircraft. A typically military fleet will consist of some civilian aircraft adapted for military purpose or used for transport, and a set of military-specific aircraft. These aircraft are able to operate in ways civilian aircraft cannot; they can use military-specific communications systems and have the ability to turn off some systems such as ADS-B. Military aircraft types that use ACARS range from modified airliners over small business jets to tankers and multi-role transport aircraft, but not fighter jets.

\vspace{-0.5cm}
\subsubsection{State}
\vspace{-0.1cm}
Aircraft used by state officials, members of governments or heads of state form a section of aviation which differs from country to country. In some states, officials are transported by the flag-carrier airline, in others this task falls to the military, and many heads of state also have their own private aircraft. For example in the United Kingdom, the Royal Family and government use state-owned, military-operated aircraft~\cite{Bourn2001}. Regardless of the operator, these aircraft are usually of a similar type to business aircraft; from small Gulfstream, Embraer and Bombardier jets to larger Airbus or Boeing jets for bigger delegations. They tend to operate in similar ways to civilian aircraft even if they are operated by the military.

\vspace{-0.5cm}
\subsubsection{Hobbyists}
\vspace{-0.1cm}
Hobbyist aviators represent the most heterogeneous group of pilots, users and aircraft. These aircraft usually operate in a visual, rather than instrument based manner meaning that flights are shorter and weather dependent. Pilot skill is likely to vary more since fewer are professionals, thus the number of flying hours significantly differ. This category includes gliders, light aircraft flown for pleasure or stunt aircraft. They are able to carry under five passengers and are not commonly used to transport said passengers from place to place. Because of the limited instrumentation, this group of aircraft does not tend to use ACARS thus we do not consider them further.





\vspace{-0.1cm}
\section{Data Sources and Collection}
\label{sec-collect}

\label{sec-access}

In order to assess the privacy breaches according to this framework, we collected ACARS data for nine months. We describe our data sources and collection approach, as well as our ethical and legal considerations

For a long time after conception, ACARS required specialist hardware to decode thus placing it out of reach to all but a determined attacker. Recently, software-defined radios such as the RTL-SDR have become available for as little as \$10, spurring the development of software enabling the decoding of previously difficult to access communications. A number of tools exist to decode ACARS, many of which are free. 

As described in Section~\ref{sec-threat}, we presume an attacker is moderately resourced and skilled in using SDRs and receiving signals. We focus on VHF and SATCOM, due to the much increased difficulties and space required to install an HF antenna. A typical VHF setup is straightforward, simply requiring an airband antenna usually available for \$50-100, fed into an RTL-SDR running ACARSDec~\cite{Leconte2015}. We collected for 242 days on the three European channels: 131.525~MHz, 131.725~MHz and 131.850~MHz. Note that we solely collect on Plain Old ACARS channels since currently, there are no publicly available tools to gather VDLm2 messages.

SATCOM uplink is located in the L-band around 1.5~GHz; reception uses a patch antenna costing \$80, fed into an RTL-SDR running JAERO.\footnote{\url{https://github.com/jontio/JAERO}} We recorded all eleven uplink channels of INMARSAT satellite 3F2 for 67 days. SATCOM downlink and HF require larger investments in antenna because of low transmission power. The former, located in the C-band around 3.5~GHz, has much shorter wavelengths than the uplink and greater path loss. On the other hand HF has much larger wavelengths thus requiring large antennas for optimal reception. 

Due to the nature of transmission over VHF and SATCOM receive one direction of communication. For VHF, we primarily receive downlink of overhead aircraft as also receiving uplink would require line-of-sight to the ground transmitter. In the case of SATCOM, uplink messages are a higher power and lower wavelength due to aircraft having limited receiver space. This is easy to intercept as the beam area is large. Ground stations receiving downlink can use bigger receivers than aircraft, allowing satellites to transmit at higher wavelength thus having a smaller beam and making it much more difficult for third parties to intercept.

Over the course of collection, we obtained \num{1651941} messages across both links, with \num{1228452} (74.4\%) being from SATCOM uplink and the remaining \num{423489} (25.6\%) being VHF. On the SATCOM link this consisted of \num{4537} aircraft, and \num{6190} on VHF with \num{1797} of these appearing on both. In Table~\ref{tab-actype} we show the number of aircraft belonging to each stakeholder group. Clearly, commercial aircraft make up the majority of all ACARS users with 74.44\%, with those qualifying as business jets comprising the other significant portion at 19.06\%. Military and state-based aircraft were observed to be much smaller at 4.91\% and 1.60\% respectively---unsurprising considering the exclusivity of these groups.

\begin{table}[t!]
\centering
\small
\caption{Breakdown by stakeholder of identifiable aircraft using the ACARS data link.}
\begin{tabular}{p{2.5cm}|p{5cm}}
Stakeholder type & Number of aircraft (\% total aircraft)\\   \hline
Private/business        & 1,701 (19.06\%)\\ 
Commercial        		& 6,645 (74.44\%)  \\ 
Military               & 438 (4.90\%)     \\ 
State-related          & 143 (1.60\%)        \\  \hline
All			          & 8,927 (100\%)        \\ \hline

\end{tabular}
\label{tab-actype}
\end{table}

\vspace{-0.2cm}
\subsection{Aircraft Metadata Sources}
\label{sources} 
There are several public data sources which provide meta-information on aircraft based on their identifiers: the aircraft registration or a unique 24 bit address provided by ICAO. This information typically includes the aircraft type (e.g., Airbus A320) and the owner/operator (e.g., British Airways), which can be exploited for further in-depth analysis and stakeholder identification. We use several public database provided by third parties for our analysis of the aircraft metadata of ACARS users, typically in SQLite or CSV format:

\begin{itemize}
\item The first database is available and constantly updated in the Planeplotter software~\cite{CentrodeObservacaoAstronomicanoAlgarve2016}. Our version of the database is from October 2016, containing 147,084 rows of aircraft data. 
\item The second database is available from Junzi Sun at TU Delft, who has been collecting information on all visible aircraft from a public flight tracker over a period of 18 months at the time of writing, amounting to 116,338 rows~\cite{Sun2016}.
\item The not-for-profit project Airframes.org is the most valuable online source for learning general aircraft information as it offers comprehensive data, including background knowledge such as pictures and historical ownership information~\cite{Kloth2016}. 
\item Lastly, for aircraft registered in the USA, the FAA provides a daily updated database of all owner records, online and for download. These naturally exclude any sensitive owner information but overall contain 320,777 records at the time of writing~\cite{FAARegistry}. 
\end{itemize}

More ACARS focussed, AVDelphi provides logs of ACARS messages for aircraft~\cite{Avdelphi}. This allows for checking of aircraft when they travel outside of our collection range.

Note that these sources are naturally noisy, since they rely on compiling many separate smaller databases and aircraft around the world are registered, de-registered and transferred regularly. However, in their combination they provide a very accurate picture of the general characteristics of the aviation environment.

\vspace{-0.2cm}
\subsection{Legal and Ethical Considerations}
Since we were aware that the likelihood of collecting sensitive data was high we upheld strong ethical conduct throughout the work. At all times data access was restricted and not disseminated in full form. Indeed, we have made great efforts to anonymize data presented in this paper such that the privacy breach cannot be exploited through our work. We ensured that all relevant laws and regulations were adhered to, and informed organizations who were leaking particularly sensitive information such as financial or medical data.

\vspace{-0.1cm}
\section{Privacy Expectations of Stakeholders}
\label{sec-expectations}

\begin{table*}[t]
\centering
\caption{Comparison of privacy requirements against whether they have been breached in ACARS usage. `N' is no breach, `V' is evidence of a breach and `X' is an explicit breach.}
\label{tab-expt}
\begin{tabular}{lllllllll}
\hline
\multirow{2}{*}{} & \multicolumn{2}{c}{Existence} & \multicolumn{2}{c}{Intention} & \multicolumn{2}{c}{Status} & \multicolumn{2}{c}{Passenger/Cargo} \\ \cline{2-9} 
                  & Required      & Breached      & Required      & Breached      & Required     & Breached    & Required            & Breached            \\ \hline
Business          & Low           & X           & High          & X           & High         & X         & High                & V                 \\ \hline
Commercial        & None          & N/A           & None          & N/A           & None         & N/A         & High                & X                 \\ \hline
Military          & High\tablefootnote{Assuming this is an operational flight, i.e. not training or similar.}          & X           & High          & X           & High         & X         & High                & V                 \\ \hline
State             & Low           & X           & High          & X           & High         & X         & High                & V                \\ \hline
\end{tabular}
\end{table*}

In this section we apply the concept of privacy expectations to the different stakeholder groups. We summarize the respective privacy requirements in Table~\ref{tab-expt} alongside the breaches, which are detailed in Section~\ref{sec-breach}. 

For each requirement we rank it as \textit{none}, \textit{low} or \textit{high}. A \textit{high} privacy expectation is critical to the stakeholder---for example, a breach might cause financial loss or personal harm. We class \textit{low} expectations as those which would be beneficial but are not placing stakeholders at risk; an example could be something which might cause inconvenience but no harm like leaking a telephone number. \textit{None} means that the information is necessarily being public. 

\begin{table}[t!]
\centering
\small
\caption{Breakdown of blocked aircraft by stakeholder. Percentages are of total aircraft in that stakeholder group.}
\begin{tabular}{p{2.5cm}|p{3.5cm}}
Stakeholder type & Number of blocked aircraft (\% aircraft for group)\\   \hline
Private/business        & 1,617 (95.06\%)\\ 
Commercial        		&  171 (2.57\%)  \\ 
Military               &  418 (95.43\%)     \\ 
State-related          &  81 (56.64\%)        \\  \hline
All          &  2,287 (25.62\%)        \\ \hline
\end{tabular}
\label{tab-acblocks}
\end{table}

In general, we base our assessment of a stakeholder's privacy requirements on the following sources:\\
\begin{itemize}
\item \textbf{Public statements:} Certain privacy requirements can be inferred directly from statements made publicly by the affected stakeholders.\\
\item \textbf{Blocking of flight data:} As explained in Section~\ref{sec-asdiblock}, schemes exist allowing stakeholders to restrict the dissemination of status, intent and existence information. Should an aircraft utilize this, we can determine that they are privacy sensitive; depending on the level of block they use, they may be sensitive to exposure of all types of information. We use the numbers provided in Table \ref{tab-acblocks} to infer such requirements for the four different groups.\\
\item \textbf{Obscuring of public registers:} Similarly to the blocking of flight data, privacy requirements (in particular existence) can be derived from stakeholders who use third-party entities to register their aircraft in order to conceal the real owner (see Section~\ref{sec-obscuring}).
\end{itemize}

\vspace{-0.5cm}
\subsection{Business}

As can be inferred from Table \ref{tab-acblocks} the majority, at over 95\%, of international business aircraft are part of schemes that block the public display of their flight data. This illustrates the overwhelming wish for protection of the intent and status of this user group. Indeed, the predecessor of the FAA ASDI scheme was originally created for business aircraft and regular statements of corresponding lobby groups on thus topic underline the severity of this privacy requirement \cite{nbaanews, faaprivacy, nbaastatement}.

Information about passengers on board business jets may be sensitive, too, and can in the case of non-hired jets typically be inferred from its owner or operator. Indeed, while the owners and operators of most aircraft globally are easily accessible in various official databases, this is not true for all jets in the business category. As discussed in Section \ref{sec-obscuring}, many individuals and corporate owners register their aircraft under other names. Instead they use hard-to-track shell companies registered overseas, or specialized services offered by trust and wealth management companies, which act as the owner or operator, illustrating a heightened need for privacy of passenger information in this category. In some cases, these tactics may also be used in an attempt to hide the existence of a specific aircraft as far as it relates to a particular owner~\cite{faaprivacy}. 

\vspace{-0.2cm}
\subsection{Commercial}

Typically, commercial aircraft have no provision to hide data on their existence, intention or status. This is partly illustrated by the fact that their records are public and typically no capability for blocks exist (apart from very few exceptions, see Table \ref{tab-acblocks}). However, as alluded to by~\cite{Roy2004}, status messages containing content relating to cargo or operational details can pose a privacy challenge for commercial aircraft. 

Furthermore, for commercial aircraft carrying passengers there are legal---and arguably ethical---requirements to ensure that any sensitive passenger data is treated in accordance with data protection legislation covered in Section~\ref{sec-dataprot}. As such we class this type of information to have a high privacy requirement for commercial aircraft.

\vspace{-0.2cm}
\subsection{Military}

Military stakeholders have a strict privacy requirement in order to carry out operations effectively. Indeed, the original conception and development of Secure ACARS stems from a military requirement to use ACARS, which due to the lack of security as standard was considered unsuitable for military traffic~\cite{Roy2000}, illustrating the high military requirement for privacy. 

Air forces, through their close government relations, are able to use all avenues to protect their privacy. For example, aircraft of the US Air Force are excluded from all public FAA records illustrating the intent to hide their existence.\footnote{Some countries' air forces, such as Switzerland with the Swiss Aircraft Register, do publish all aircraft records in full~\cite{FederalOfficeofCivilAviationFOCA2017}.} On top of this, military aircraft typically request to have their flight information blocked from public feeds; we observed that more than 95\% of all military ACARS users apply such blocks (see Table \ref{tab-acblocks}). This demonstrates a high requirement to hide not only existence, but also status and intention, thus concealing sensitive information which may reveal operational movements. 

As far as military aircraft are used to transfer military or government leaders, their passenger information may also be considered highly sensitive. The same applies for the identity and personal information of any passenger in an aircraft operated by the military as stated for example by the US Air Force~\cite{Roy2000,Adams2006}.

\vspace{-0.2cm}
\subsection{State}

As military-operated aircraft prove to be one common approach for this category, we consider them to have a similar set of privacy requirements. Many state aircraft are blocked from public feeds and also considered of sensitive nature in public aircraft records. A related non-military example would be surveillance aircraft used by the FBI, which are registered to a number of front companies and excluded from the FAA records \cite{strohmeier2016cycon}. 

Some requirements may differ, however. Depending on the operating country, existence and intention are potentially less sensitive since many state movements are publicly announced and government aircraft are part of all official records. In other countries, state and government aircraft have been involved in several scandals with regards to their movements and sometimes simply the very nature of their existence \cite{fr24dictators}. These differing approaches are reflected in Table \ref{tab-acblocks}, where only slightly more than half of all state and government aircraft are blocked from public display. In any case, due to the high-profile nature of those on board, passenger information relating to these aircraft is likely to be extremely sensitive.

\vspace{-0.1cm}
\section{Privacy Breaches}
\label{sec-breach}

With stakeholders matched to privacy expectations, we use this section to demonstrate that ACARS is a vehicle for systematic leakage of sensitive data. Due to the fact that we only observe part of the bidirectional link for both SATCOM and VHF we define breaches as `\textit{no evidence}', `\textit{evidence}' and `\textit{explicit}'. Evidence of a breach is where messages are seen on  an observed link which would incite a leak on the other part of the link---for example, requesting personal information of a passenger. An explicit breach is where we observe instances of sensitive information being leaked on the observed link. We summarize the results of this section along with the privacy requirements defined in Section~\ref{sec-expectations} in Table~\ref{tab-expt}, noting that for all cases, our observed links either provided evidence of leaks or more commonly, explicitly leaked sensitive information.

\vspace{-0.2cm}
\subsection{Business}

Even with business leaders indicating significant privacy requirements, this category of aircraft leaks a vast amount of information. More specifically, every privacy requirement is compromised in the course of using ACARS, either due to it being in the clear or by using weak attempts at encryption. 
We summarize the privacy breaches discussed in this section through Table~\ref{tab-biz-vhf} for VHF and Table~\ref{tab-biz-sat} for SATCOM.

Although a `low' requirement, a cursory survey of US-registered aircraft observed allowed us to breach the existence requirement by looking them up via public FAA registers as described in Section~\ref{sec-obscuring}. On the VHF we saw 750 blocked aircraft (of 1002 observed, 75.0\%) registered to shell companies, with SATCOM seeing 615 (of 881, 70.0\%). We do believe that with further manual work, details on the obscured business aircraft can be obtained.

\begin{table}[t]
\centering
\caption{Summary of privacy breaching VHF ACARS usage by business aircraft (A/C). Percentages are of total number of business aircraft and  messages from business aircraft respectively.}
\label{tab-biz-vhf}
\begin{tabular}{p{1.3cm}|p{1cm}p{1.2cm}p{1cm}p{1.8cm}}
\hline
                   & Aircraft     & Blocked A/C & Messages      & Msg. from Blocked A/C \\ \hline
Position Reports   & 541 (51.3\%) & 522 (49.5\%)     & 3,859 (33.0\%) & 3,689 (31.5\%)                  \\ \hline
Clearance/ ATIS     & 168 (15.9\%) & 159 (15.1\%)     & 1,055 (9.0\%)  & 431 (3.7\%)                    \\ \hline
Encrypted Messages & 47 (4.5\%)   & 47 (4.5\%)       & 843 (7.2\%)   & 843 (7.2\%)                    \\ \hline
\end{tabular}
\end{table}

Primarily on the VHF link, some business aircraft transmit positional reports, with some containing more than just position alone. At the very least, they contravene the status privacy requirement. We observed that 541 (51.3\%) aircraft transmit positional reports, with 522 (49.5\% of all, 96.5\% of position reporting aircraft) having ASDI blocks. Indeed, 3689 (95.6\% of position reporting) messages came from blocked aircraft, undermining their effort to block their actions.
\begin{table}[]
\centering
\caption{Summary of privacy breaching SATCOM ACARS usage by business aircraft (A/C). Percentages are of total number of business aircraft and  messages from business aircraft respectively.}
\label{tab-biz-sat}
\begin{tabular}{p{1.3cm}|p{1cm}p{1.2cm}p{1cm}p{1.8cm}}
\hline
                   & Aircraft     & Blocked A/C & Messages   & Msg. from Blocked A/C \\ \hline
Clearance          & 38 (4.1\%)   & 38 (4.1\%)       & 72 (0.3\%)   & 72 (0.3\%)                     \\ \hline
ATIS               & 185 (19.7\%) & 172 (18.3\%)     & 693 (3.2\%)  & 655 (3.0\%)                    \\ \hline
Encrypted Messages & 131 (13.8\%) & 124 (13.2\%)     & 1,703 (7.9\%) & 1,598 (7.4\%)                   \\ \hline
\end{tabular}
\end{table}

Business aircraft regularly reveal intention by partaking in ATC clearance exchanges via ACARS. The ability to receive full intention information relies on geographical position---if a receiver is near to an ATC center they will receive destinations and routes. Whilst our collection location was not located in such a position, we could instead observe other parts of clearance exchanges indicating that business aircraft did use this approach. Additionally, aircraft can request information from the Automatic Terminal Information Service (ATIS) which announces information relevant to arrival or departure at an aerodrome. Since an aircraft has to request this under the ACARS label `B9' (see~\cite{AeronauticalRadioInc.ARINC2014} for further details) it is done with purpose and reveals the destination aerodrome of the aircraft thus revealing intention data. 

On the VHF link, we observed 168 (15.9\% of all) aircraft involved in clearance or ATIS exchanges with 159 (15.1\%) having ASDI blocks. These aircraft transmitted 1055 messages, with 431 (40.9\%) being from blocked aircraft in doing so counteracting their privacy efforts. On the SATCOM link, fewer clearance messages exist---only 70 messages from 38 aircraft, all of which are blocked. However, 185 (19.7\%) aircraft use ATIS over this link, 172 (18.3\%) of which are blocked. These aircraft transmitted 693 messages of this type, 655 (94.5\%) of which were from blocked aircraft thus revealing their destination aircraft. Clearly a significant amount of intention information is revealed thus is an explicit breach.

\begin{figure}
\centering
\includegraphics[scale=0.35]{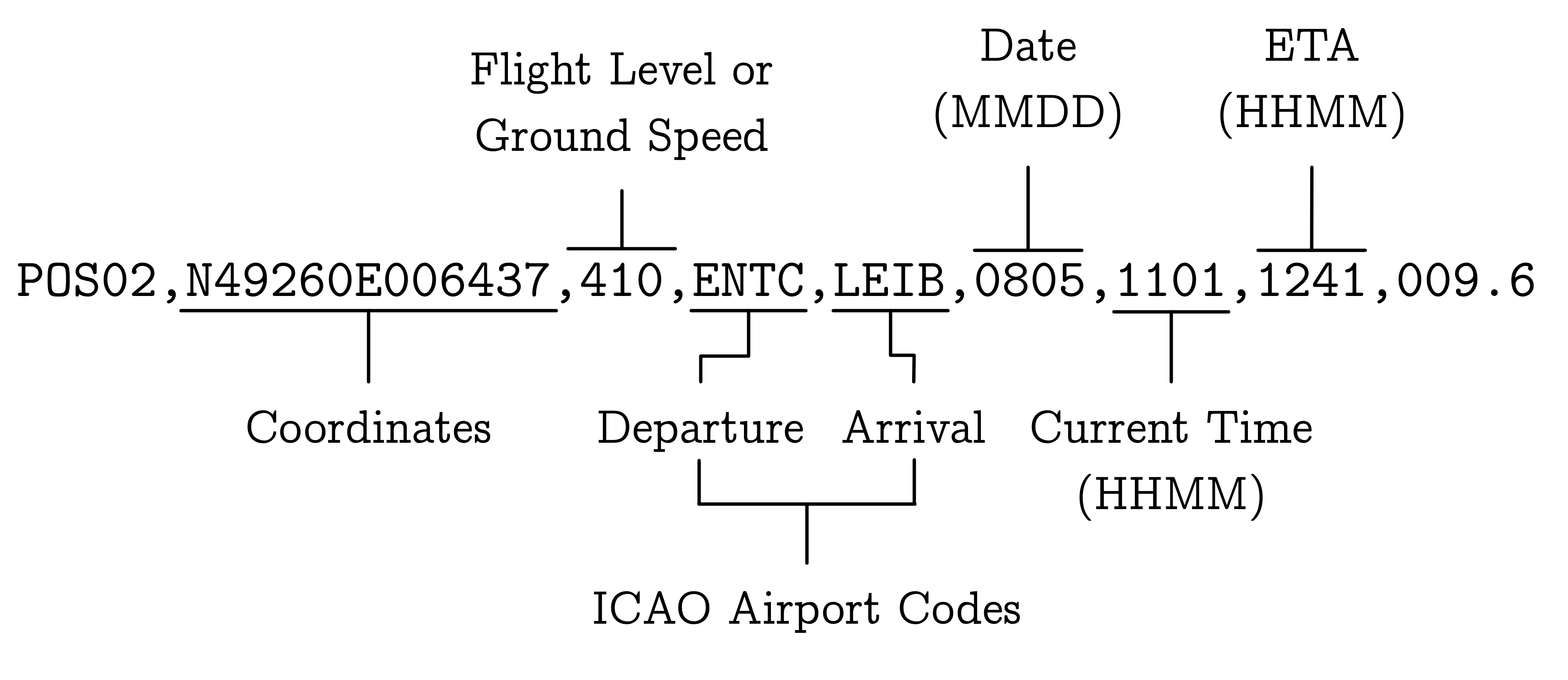}
\caption{Labelled plaintext message format of encrypted positional reports, from~\cite{Smith2017}.}
\label{fig-ct-msg}
\end{figure}

As explained in Section~\ref{sec-security-mechanisms}, attempts at using encryption to protect intention and status by some types of business aircraft fail on account of the cryptography being weak. An example position report, which as described in~\cite{Smith2017} accounted for approximately a third of messages across VHF and SATCOM, is shown in Figure~\ref{fig-ct-msg}. We observed that all 47 aircraft transmitting 843 encrypted messages on VHF were ASDI blocked. On SATCOM, 131 (13.8\%) aircraft transmitted encrypted messages, with 124 (13.2\%) being blocked---the latter group accounted for 1,598 (93.8\%) of the encrypted messages over this link.

\begin{figure}
\texttt{\small
(2AAFB12-JAN-17 14: 55Z) subjet: \\
EMAIL FROM: DISPATCH AT \textbf{[REDACTED]}.COM\\
HELLO. BROKERS SAY PAX WAS PICKED UP in central Copenhagen}
\caption{Extract from a free-text SATCOM message, discussion where the passenger has been collected from}
\label{fig-broker-mesg}
\end{figure}

As with commercial aircraft passenger information is much more anecdotal due to much of it being transferred in the free-text format, however here we did not observe any significant explicit breaches, instead evidence of breaches on the other half of the link. SATCOM is used to transfer emails, revealing 47 email addresses---some to aviation businesses, though some to Hotmail and Yahoo addresses. It is worth noting that these messages are simply uplink so do not always contain email content---collecting downlink would provide more email body data. From the email bodies available, we can discern that these aircraft are transmitting passenger information over plaintext links. For example, from one blocked aircraft with unobscured registration details, we observed the message content as shown in Figure~\ref{fig-broker-mesg} (translated after collection). This indicates part of a discussion about a passenger location thus provides evidence of a breach.

\vspace{-0.2cm}
\subsection{Commercial}

As indicated in Table~\ref{tab-expt}, commercial aircraft are considered to have the lowest privacy requirements of all stakeholder groups since a lot of information relating to their flights is made public. However, the leaks from this type of aircraft are arguably the most significant and explicit, especially for passengers.

Most directly impacting is the transmission of credit card details, observed in use by two airlines. Whilst not extremely common, this occurs with sufficient regularity and message structure to indicate that this is part of crew procedure. We saw three levels of detail on these transmissions, from most to least sensitive:
\begin{itemize}
\item Full details, including a credit card number, CVV number, expiry date, amount and sometimes name.
\item Partial details, including a partial credit card number, authorization code or less sensitive details such as expiry date or amount.
\item Transactional context, such as whether a given transaction (per authorization code) has been authorized, denied, or otherwise.
\end{itemize}
Since SATCOM collection focussed on uplink messages, most were relating to authorization with some references to value. VHF, however, contained six full sets of credit card details with five providing transaction values; for each, the value was over \$500. Both SATCOM and VHF contained partial details, three and four sets respectively, and SATCOM carried 11 context messages. This is a significant breach which not only contravenes data protection legislation but also Requirement 4 of the Payment Card Industry Data Security Standard (PCI-DSS)~\cite{PCISecurityStandardsCouncil2016}. This requirement handles the transmission of data over networks to which a malicious user can easily gain access, which is the case with ACARS. Breaching these regulations can see an organization have its ability to handle card payments withdrawn.

Free text messages are regularly used to allow crew and ground bases to communicate about passenger issues, often transmitting passenger information, leaking it due to the clear text nature of ACARS. During collection we received \num{2811} messages relating to passengers.

ACARS lends itself to fast communication ahead to destination airports so is used to deal with medical emergencies. On SATCOM uplink, ground stations would often ask for details on unwell passengers providing evidence of downlink breaches. Over both links we observed 28 messages containing medical information. Within this, we observed nine instances each of messages containing passenger names and further medical information such as condition details, with 21 instances of medical context (i.e. non-medical data or names relating to an emergency, for example seat numbers or medics waiting on landing), providing evidence of breaches on the other part of the link. Clearly, in the instances where passenger name and medical information are transmitted via ACARS, data protection legislation is clearly breached.


In order to inform passengers ahead of landing about onwards journeys, some crew contact airline operations centers to request connecting flight status. However, some airlines transmit passenger names and onward destinations as part of this. We observed 30 messages transmitted on SATCOM containing 236 passenger names.  Of these, 15 were from one airline which, from three aircraft leaked 210 passenger names, seat numbers and corresponding connection details. It appears that for this airline, any passenger with a connection will have their data transmitted. In the case of the less serious breaches (i.e. the remaining 15 messages), 26 passenger names were transmitted and for 18 of these passengers, an onward destination was also observed. 

Whilst this type of breach is not as serious as the cases above, it does reveal a lot of information about passenger movements thus is an explicit privacy breach. It is worth noting that the German airline Lufthansa transmitted 250 messages apparently related to connection information over VHF and SATCOM; all of these messages were encrypted, demonstrating that they consider it sufficiently important information to be worth protecting.


\vspace{-0.2cm}
\subsection{Military}

\begin{figure}
	\centering
	\includegraphics[width=\columnwidth,trim=1.5cm 4cm 1cm 5cm, clip]{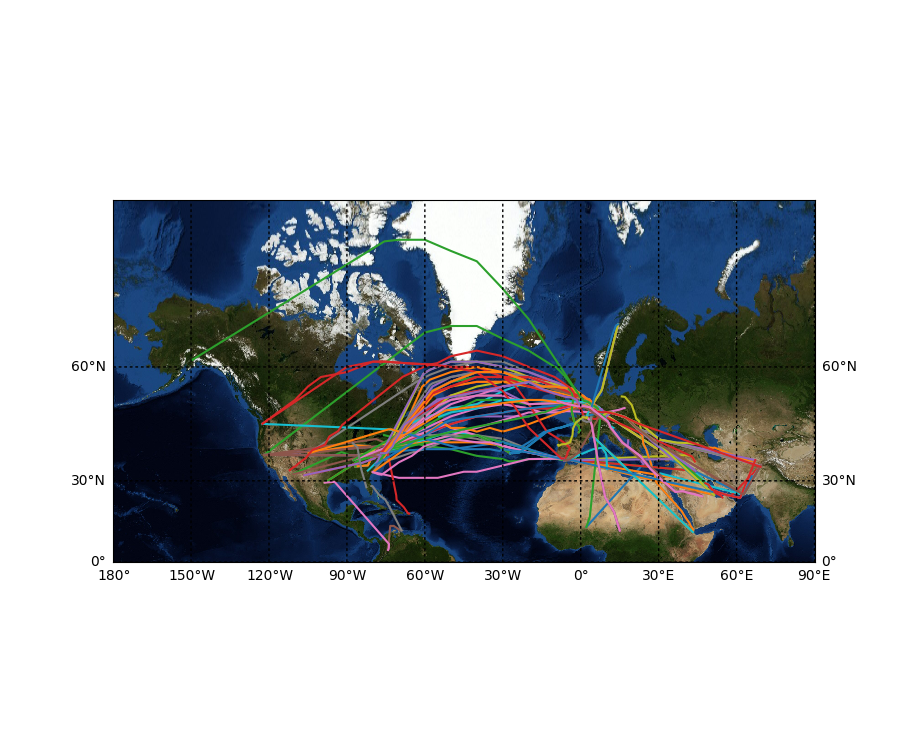}
	\caption{Plotted flight tracks for all flight plans received for military aircraft. Colours were assigned randomly and do not reflect multiple flight from the same aircraft.}
	\label{fig-mil-tracks}
\end{figure}

\begin{table}[]
\centering
\caption{Military aircraft and messages leaking existence and intention information. Percentages are of all aircraft/messages from the same category vulnerable to privacy breaches. Message percentages calculated over all, including non-ACARS messages, for existence and of ACARS-only messages for other categories.}
\label{tab-mil-stats}
\begin{tabular}{l|rr }
	Leak type & \vtop{\hbox{\strut Number of aircraft}\hbox{\strut (\% of aircraft)}} & \vtop{\hbox{\strut Number of messages}\hbox{\strut (\% of messages)}} \\ \hline
	Existence &    53 (12.5\%)                       & 120,100 (82.8\%)                       \\ \hline \hline
	Flight Plans &  287 (67.9\%)                     & 1,260 (5.1\%)                           \\
	Clearance \& ATIS &   114 (27.0\%)               & 485 (1.9\%)                        \\
	Weather     &       227 (53.7\%)                 & 1,792 (7.2\%)                        \\ \hline
\end{tabular}
\end{table}

The number of military aircraft observed is typically lower than the other categories. A summary of the most important numbers can be found in Table~\ref{tab-mil-stats}.  On the VHF link, less than one percent of all observed aircraft could be categorised as belonging to military forces, compared to slightly more than $9\%$ on the satellite link. Of those seen via satellite link, $12.5\%$ did not receive any ACARS messages but only non-ACARS data, such as network management information. This leaks existence even with no active communication over ACARS. Furthermore, the vast majority (95\%) of military aircraft seen use ASDI blocks thus have their existence privacy requirement breached.

Over our collection phase, we received over \num{1000} flight plans being transmitted to more than $250$ military aircraft over the satellite link. We were able to reconstruct the flight path for most of them, as shown in Figure~\ref{fig-mil-tracks}. This breaches the intention requirement considerably. We also received close to \num{1800} weather reports sent to crew of more than $200$ military aircraft. While not leaking the exact route, these messages specify departure and arrival airports and sometimes navigation waypoints.

During one and a half months of satellite link collection we spotted a handful of messages of a military aircraft fleet commonly used to transport the head of state of a western government. These messages breached privacy in various ways. First of all, we received flight plans which we were able to verify using the OpenSky Network's (described in~\cite{Schafer2014}) ability to receive both Mode S and ADS-B transmissions, when the aircraft crossed the reception range of the sensor network. Additionally, we received status messages indicating passenger occupancy of a `state room' and number of senior staff and general passengers on board. Amongst the received data were cargo lists, indicating presence and absence of objects called \emph{white elephant} and \emph{grey ghost}. As such this indicates the existence of a breach of passenger information but not necessarily an explicit breach, due the the limited detail of the data transmitted.

Military aircraft have the ability to turn ADS-B off in order to provide privacy for existence---the only category of aircraft in our framework which have a definite need to hide the fact that they are in the air. It is not enough to simply turn off ADS-B reporting in order to hide ones existence, however. From our data, we were able to reconstruct that $21\%$ and $48\%$ of military aircraft transmitting over satellite and VHF, were not leaking more than their existence, while $79\%$ and $52\%$ were leaking information for at least in one of the remaining privacy categories highlighted in Table~\ref{tab-expt}.

\vspace{-0.2cm}
\subsection{State}
\begin{table}[]
\centering
\caption{State aircraft and messages leaking existence and intention information. Percentages are of all aircraft/messages from the same category vulnerable to privacy breaches. Message percentages calculated over all, including non-ACARS messages, for existence and of ACARS-only messages for other categories.}
\label{tab-state-gov-stats}
\begin{tabular}{l|rr }
	Leak type & \vtop{\hbox{\strut Number of aircraft}\hbox{\strut (\% of aircraft)}} & \vtop{\hbox{\strut Number of messages}\hbox{\strut (\% of messages)}} \\ \hline
	Existence &    46 (31.8\%)                       &    16,026  (84.8\%)                       \\ \hline \hline
	Flight Plans &  6 ( 5.5\%)                        &    19  (0.7\%)                           \\
	Clearance \& ATIS &    22 (20.0\%)               &    108  (3.8\%)                        \\
	Weather     &       31 (28.2\%)                  &    234 (8.2\%)                        \\ \hline
\end{tabular}
\end{table}

Governmental and state aircraft form the smallest proportion of our data, at $1.6\%$ of received messages. However, SATCOM proved adept in receiving these messages, with state and military aircraft combined comprising over $10\%$ of aircraft seen, due to its global reach.  

While existence is only classified as a `low' requirement in general for this stakeholder group, individual flight missions might still require the possibility to use radio silence. A total of $42\%$ of all state and government aircraft were receiving satellite communication, however they did not receive any ACARS messages. Over one third of all governmental aircraft could therefore potentially gain privacy by not communicating over the satellite link as shown in Table~\ref{tab-state-gov-stats}. We can see that ACARS also breaches existence privacy for the 56\% of state aircraft using ASDI blocks through their use of the system.

\begin{figure}
\centering
\includegraphics[scale=0.2]{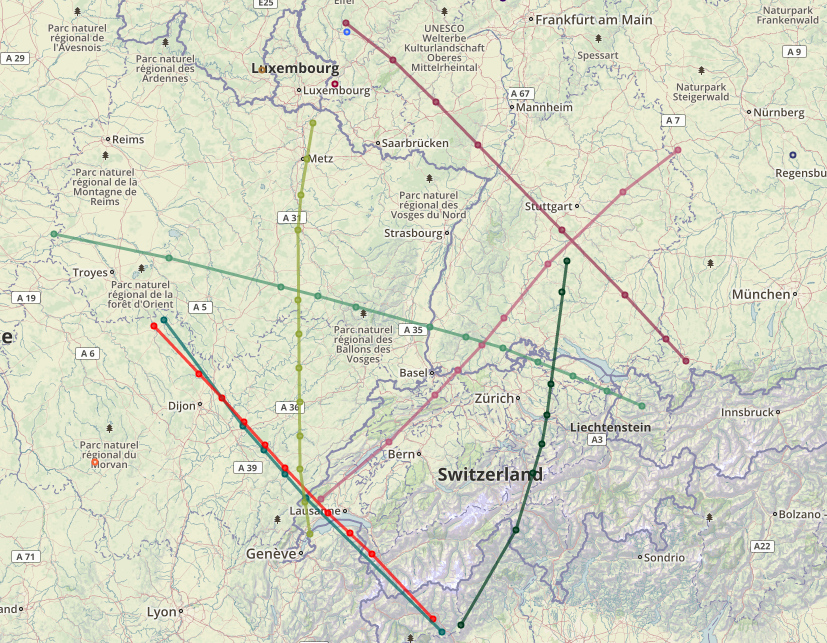}
\caption{Plotted location reports from a government aircraft transmitting positional reports.  Where there were multiple position reports in a short time period, these are joined together as a track.}
\label{fig-gov-tracks}
\end{figure}

Aircraft used on behalf of one government regularly transmitted positional claims over the course of the collection period; for one of these aircraft, Figure~\ref{fig-gov-tracks} displays plotted reports joined up into tracks where messages were sent in over a short time period. Note that this aircraft has no flights available to view on Flightradar24, thus by transmitting these reports over ACARS, they are explicitly breaching privacy and undermining other efforts.

Another aircraft from a European government's fleet regularly received ACARS messages over satellite link indicating that data link position reporting had been turned off. Looking up this aircraft on AvDelphi indicated that position reports were received nearly every day with the exception of days where `position reporting turned off' messages had been received via satellite. The same aircraft was also seen transmitting position reports on VHF when it passed through our reception range.

We found communication directly related to state aircraft intention in either clearance and ATIS messages or in weather reports. As Table~\ref{tab-state-gov-stats} shows, one fifth of all government related aircraft received clearance and ATIS messages containing destination airport codes. Slightly more than $25\%$ of those receiving weather reports were also leaking the destination airport. We received a small number of weak encrypted messages as outlined in Section~\ref{sec-security-mechanisms}, used by business jets. These messages contained weather reports leaking departure and destination airports thus adding to the breach of intention information.

A small number of governmental aircraft were also observed transmitting clear text e-mail messages via the ACARS satellite link. The nature of these messages was mainly flight status related, however, additionally leaking names and e-mail addresses belonging to fleet operator or government employees. However since email address are of limited sensitivity and we saw no evidence of further passenger data, we do class this as evidence of breach of passenger information requirement rather than an explicit breach.

\vspace{-0.1cm}
\section{Implications of Breaches}
\label{sec-discussion}

Clearly, the ACARS data link in its current form poses a significant privacy challenge; leaks are widespread and consistent enough to constitute a serious and long-running issue. In this section, we discuss the key implications of the observed privacy breaches and provide recommendations to prevent or reduce further data leakage.

As we have shown, many aviation stakeholders already go to great lengths to protect their respective privacy interests but are regularly undermined by their use of ACARS. Indeed, for each group, we have observed leaks with the potential to cause serious harm, either politically, reputationally, economically or otherwise.

For example, breaches relating to passenger information put the ACARS user at risk of financial and reputational loss, by breaking relevant data protection laws. Overall, we note a major lack of awareness and complacency around the modern use of ACARS as a data link. While some individuals within aviation have warned against its imprudent and thoughtless use \cite{Wolper1998,aeropers}, most ACARS users seem to be either unaware of its security problems or use it all the same.

\vspace{-0.2cm}
\subsection*{Mitigation and Countermeasures}
Many of the leaks described in Section~\ref{sec-breach} are as a result of the message content; without it, the most a single receiver can determine is the aircraft existence. Thus, protecting message content is the most effective way to address leaks. There are two ways of achieving this with ACARS deployed as it currently is: educate users to avoid transferring sensitive data via ACARS, or deploy an ACARS Message Security solution to all relevant aircraft.

Deploying AMS would be the most complete and effective countermeasure, as it can provide message authentication, confidentiality and integrity. This would effectively solve the issue of leaks to a passive observer but does come with challenges. As with any distributed security solution, implementing a public-key infrastructure is costly and requires thoughtful, security-conscious design. Especially in the case of aircraft, which must be able to communicate with unexpected ground stations, keeping up-to-date credentials for all communication partners is a challenge. Currently the only option is Secure ACARS~\cite{Roy2004}, an implementation of AMS that comes at a surcharge to the ACARS service. It could be that this cost factor is key factor in its almost non-existent deployment, even though the cost of the investment is likely offset by the potential reputational damage and legal costs.

While the temptation of cheap, proprietary cryptography-based solutions is great (as observed in use on some business jets~\cite{Smith2017}), weak encryption is to be avoided at all costs. Providing the illusion of security and no more, this approach detracts from the importance of quickly deploying well-developed solutions to aircraft.

The second option lies in the education of ACARS users such that they transfer sensitive information in another manner. Of course, this is not mutually exclusive to using encryption, but is more likely to be deployable in a timely fashion. It would require ACARS users to identify the sensitive information transferred and arrange other methods of transfer, if needed. A key challenge here lies in continuing to provide services via ACARS where sensitive information is crucial. 


In the longer term, steps should be taken to move away from ACARS completely. Since it was designed with a significantly weaker threat model in mind---i.e. one of no malicious activity---it is not equipped to deal with cybersecurity threats. As discussed, uptake on available security solutions has been limited, which indicates that a newly developed data link with security as the default, may be the better option. However, given typical technological cycles in aviation, this would take decades to deploy fully \cite{strohmeier2016cycon}.

\vspace{-0.1cm}
\section{Related Work}
\label{sec-related}

Nearly a decade before avionics communication gained interest in the scientific community, the United States Air Force published concerns about the security and privacy of ACARS~\cite{Roy2000}. To keep military frequencies clear for tactical communication, they propose an encryption and authentication system to allow military aircraft to communication over commercial data link.

Since this, work has highlighted the privacy and security issues of ACARS. Most recently,~\cite{Smith2016} highlighted some anecdotal evidence for these issues indeed motivating the more comprehensive approach taken in this paper. In~\cite{Storck2013}, the impact of ACARS being transmitted in the clear is discussed specifically with respect to the level of trust which can be placed in the link, though without further quantification. Outside of the academic community, ACARS has received some attention at hacking conferences due to its lack of integrity and authentication mechanisms~\cite{Teso2013}.

A survey in the avionics community was conducted in~\cite{Strohmeier2016a} to find how the actual users assess the security and trustworthiness of the avionics protocols. Most responders believed the likelihood of attacks to be low and trustworthiness was rated above average for most protocols, including ACARS.

Privacy properties of the avionic surveillance system ADS-B are investigated in~\cite{Sampigethaya2011,Sampigethaya2013}. Since a passive attacker can trivially receive ADS-B messages, they investigate the effectiveness of the standardized privacy approach---identifier randomisation. They show that knowledge of one identifier allows calculation of subsequent identifiers for the same aircraft. Consequently, they propose the usage of decorrelated random identifiers to mitigate this.


Passenger privacy is not a purely aviation focussed problem and does not restrict itself to public transport either. In~\cite{Rouf2010}, the authors investigate the privacy and security of current car sensor systems focussing on tire pressure systems. An attacker is able to read the static identifiers of the tire pressure sensors from tens of meters away thus allowing tracking of individual vehicles crossing the an attacker checkpoint. Exploting in-car networks~\cite{Kleberger2011}, discusses how an attacker can use the CAN bus to extract data from the vehicle. Indeed, most of these systems lack even basic security features.


\vspace{-0.1cm}
\section{Conclusion}
\label{sec-conclude}

In this paper, we analyze the privacy breaches arising from ACARS usage. This is an important avionic data-link system, used for both ATC and fleet management purposes by a significant number of aircraft. Our results demonstrate significant breaches in normal usage of the system. For example, we observed non-commercial aircraft undermining their efforts to limit public information on their flights by openly transmitting with ACARS. In the commercial case we saw transmission of very sensitive data such as credit card details and medical information. In order to evaluate these breaches we propose a framework which identifies aviation stakeholder groups and ties them to privacy concepts. Using this we show that every instance of a privacy requirement is either breached or has strong evidence of such a breach, for each stakeholder.

Based on our findings, we highlight that ACARS alone is not responsible for leaking information---instead, it is a combination of lacking a secure default and misuse by users. We propose that in order to address this, stakeholders need to push for use of ACARS Message Security implementations such as Secure ACARS and in cases where they are operating over an unsecured link, be wary of the sensitivity of the information they are transmitting. In the longer term, however, ACARS should be replaced with a data link better suited to a modern threat model.

\bibliography{biblio,refs}
\end{document}